# PQC: GENERALIZED ELGAMAL CIPHER OVER *GL(8, $F_{251}$)*

P. Hecht[1]

***Abstract*— Post-Quantum Cryptography (PQC) attempts to find cryptographic protocols resistant to attacks by means of for instance Shor's polynomial time algorithm for numerical field problems like integer factorization (IFP) or the discrete logarithm (DLP). Other aspects are the backdoors discovered in deterministic random generators or recent advances in solving some instances of DLP. The use of alternative algebraic structures like non-commutative or non-associative partial groupoids, magmas, monoids, semigroups, quasigroups or groups, are valid choices for these new kinds of protocols. In this paper, we focus in an asymmetric cipher based on a generalized ElGamal non-arbitrated protocol using a non-commutative general linear group. The developed protocol forces a hard subgroup membership search problem into a non-commutative structure. The protocol involves at first a generalized Diffie-Hellman key interchange and further on the private and public parameters are recursively updated each time a new cipher session is launched. Security is based on a hard variation of the Generalized Symmetric Decomposition Problem (GSDP). Working with GL(8, $F_{251}$) a 64-bits security is achieved, and if GL(16, $F_{251}$) is chosen, the security rises to 127-bits. An appealing feature is that there is no need for big number libraries as all arithmetic if performed in $\mathbb{Z}_{251}$ and therefore the new protocol is particularly useful for computational platforms with very limited capabilities like smartphones or smartcards.**

## 1. INTRODUCTION

Post-Quantum Cryptography (PQC) is a relatively new cryptologic trend that recently acquired an official NIST status [1, 2] and which aims to be resistant to quantum computers attacks (like Shor algorithm). But PQC not only cover against that menace, it works also as a response against side-channel attacks [3], the increasing concern about pseudo-prime generator backdoor attacks (i.e. Dual_EC_DRBG NSA [4]) or the development of quasi-polynomial discrete logarithm attacks [5] which impact severely against current de facto standards [6] of asymmetric cryptography whose security rest on integer-factorization (IFP) and discrete-logarithm (DLP) over numeric fields. And more, sub-exponential time complexity attacks on many instances appear [5][6]. Shor algorithm [7] opened a quantum computing way to break current asymmetric protocols. As a response, there rise an increasing interest in some simple solutions like Lattice-based, Pairing-based, Multi Quadratic, Code-based, Hash-based, Non-Commutative and Non-Associative algebraic cryptography [1, 2, 8 to 13].

A whole branch of new protocols was developed which do not rely on extended precision arithmetic's and instead exploit internal asymmetry of abstract algebraic structures like partial grupoids, categories, magmas, monoids, quasigroups, groups, rings, loops or neofields [9 to 24]. The new developed one-way trapdoor functions (OWTF) include conjugator search (CSP), decomposition (DP), commutative subgroup search (CSSP), symmetric decomposition (SDP) and generalized symmetric decomposition (GSDP) [9, 15, 17, 25, 26].

This paper focus a simple solution using the general linear multiplicative subgroup over prime field $F_{251}$, represented as $GL(d, F_{251})$, $d$ is the square matrix order. All arithmetic operations are into $Z_{251}$. The prime characteristic 251 is the biggest one fitting into a byte. As advantage, no big number libraries are involved, memory requirement reduced and fast computation expected. As a necessary condition for asymmetric cryptography, a hidden commutative subgroup is developed inside. PQC studies were purposely followed by the author over his past research [27 to 32].

## 2. ALGEBRAIC CONCEPTS

Let $p$ be a prime, $d$ any integer >1, $q=p^d$ and $F_p[x]$ the polynomial extension of the prime field $F_p$. The number of square matrices of order $d$ and values in $F_p$ is $p^{d^2}$, and of those $p^{d^2-d}$ are nilpotent [33 to 36]. The number of elements in the general linear group of $d$-order non-singular square matrices is:

$$\left|GL(d, F_p)\right| = \prod_{i=0}^{d-1}(p^d - p^i) \qquad (1)$$

A non-singular matrix or $d$-order whose monic characteristic polynomial is irreducible in $F_p$, generates a cyclic (thus commutative) subgroup $P_d$ of $M_d = GL(d, F_p)$. Each $d$-degree irreducible polynomial $f(x)$ in $F_p[x]$ field has a square companion matrix of $d$-order who acts as a generator of the multiplicative cyclic subgroup $P_d$, and each member of this subgroup corresponds to a unique monic characteristic polynomial of at most $d$-$1$ degree [ ]. The $N_{tot}$ number of non-trivial (null or unitary) monic $d$-degree $f(x)$ over $F_{251}$ field is:

$$N_{tot} = p^d - 2 \qquad (2)$$

Using Möbius $\mu$ function, the $N_p(d)$ number of monic irreducible $d$-degree polynomials over $F_p[x]$ field is:

---

[1] Pedro Hecht: Maestría en Seguridad Informática, FCE-FCEyN-FI (Universidad de Bs Aires) phecht@dc.uba.ar

$$N_p(d) = \frac{1}{d}\sum_{r|d} \mu(d) p^{d/r} = \frac{p^d - 2}{d} = \frac{N_{tot}}{d} \qquad (3)$$

To generate a random $d$-order monic irreducible polynomial over $F_p[x]$, we use the probabilistic Algorithm 4.70 [6] whose complexity is $O(m^3 (\lg m)(\lg p))$ and requires approximately $d$-trials. Once found, it is translated into the companion matrix [33]. Uppon, it is of interest to find its order, because that would be the number of elements of the commutative subgroup $P_d$ of the $M_d$ matrix group. Whatever this value is, it must be a divisor of the multiplicative subfield order ($= p^d - 1$) and if it were maximal, the irreducible polynomial would be a primitive one. To calculate polynomial orders, a modified version of Algorithm 4.77 [6] can be used.

Clearly using an irreducible polynomial in an extension field is a method of generating a $P_d$ commutative subgroup of the non-singular modular square matrices, but there exists another way to achieve the same goal. For matrices, the necessary and sufficient condition for two symmetric (diagonalizable) matrices to commute, is that they share the same orthonormal basis, that means the same eigenvectors $P$ matrix [34, 35]. If we start from two different diagonal matrices $D_1$, $D_2$ ; then the transformed $A$ $(=P D_1 P^{-1})$ and $B$ $(=P D_2 P^{-1})$ commute $(AB = BA)$. The later approach is computational faster than the first one, so it will be followed in our protocol.

## 3. Cryptographic Aspects

Security of an asymmetric cipher protocol always relies on a hard OWTF [6]. Here we propose a generalized ElGamal cipher selecting GSDP as the one-way trapdoor function. If the algebraic structure and OWTF are well selected, a provably secure protocol could be developed [9, 15]. This sounds good, but it is not easy to prove such a claim [37]; so caution at use is strongly advised. In our case, the GSDP could be stated as follows

$$\text{Let } G \text{ a non commutative group and } S \text{ a commutative subgroup, given } (x, y) \in G^2 \text{ and } (m, n) \in \mathbb{Z}, \text{ find } z \in S \mid y = z^m\, x\, z^n \qquad (4)$$

This structure resembles a generalized discrete logarithm (GDLP) or a conjugation search problem (CSP). GSDP is more difficult as the first one, as no numeric field is directly involved and because the vectorially structure of elements involved. GSDP is clearly a generalization of CSP, so a harder solution must be expected. GSDP is supposed to be one of the hardest challenges in group theory [9, 14, 15, 16, 17]. As no cryptanalytic quantum algorithm is on sight and probably does not exist, the present protocol belongs to the PQC set. Of course, this statement should be proven, a question beyond the purpose of this paper.

In our protocol, we use a harder variety of GSDP, with less known information. We call it blind general symmetric decomposition problem (BGSDP), and it states as

$$\text{Let } G \text{ a non commutative group and } S \text{ a commutative subgroup, given } y \in G \text{ but unknown } [\, x \in G, (m, n) \in \mathbb{Z}],\ \text{ find } z \in S \mid y = z^m\, x\, z^n \qquad (5)$$

Not only this kind of generalized discrete logarithm problem is at least difficult as GSDP, in our case we change all hidden parameters each time a new cipher session is started. We accomplish this with an iterated update of those parameters.

## 4. Cipher Protocol

In our version, we work with two entities (Alice and Bob), but this could be easily generalized for any number of participants. All arithmetic operations should be assumed belonging to field $F_{251}$. The setup steps (Table 2.) involve a generalized Diffie-Hellman key exchange. At following box, common symbols are explained as used along this protocol.

TABLE I
SYMBOLS AND DEFINITIONS.

| |
|---|
| $\in$ – *belongs to* |
| $\in_R$ –*randomly selected element in* |
| $\forall\neq$ -*all different elements* |
| $M_8 \equiv GL(8, F_{251})$– *non-commutative group* |
| $P_8 \in M_8$ – *commutative subgroup* |
| $D_A$ , $D_B$ – *diagonal matrices* |
| $K^{\nearrow}_{8,1}$ –*left upward first non-zero term of the secondary diagonal* |
| $K^{\swarrow}_{1,8}$ –*right downward first non-zero term of the secondary diagonal* |
| $K^{\searrow}_{1,1}$ – *left downward first non-zero term of the principal diagonal* |
| $K^{\nwarrow}_{8,8}$ – *right upward first non-zero term of the principal diagonal* |
| $\Longrightarrow$ *send publicly to the other entity* |
| *validation* – *greyed consistency proof* |

TABLE II
SETUP STEPS

| | | |
|---|---|---|
| Any entity begins | $P \in_R M_8 \Longrightarrow$<br>$G \in_R M_8 \Longrightarrow$ | |
| | ALICE | BOB |
| Generating private elements | $k_1, k_2 \in_R {\mathbb{Z}^*_{251}}^2$<br>$\forall\neq \lambda_1 \dots \lambda_8 \in_R \mathbb{Z}^*_{251}$<br>$D_A = (\lambda_1 \dots \lambda_8)$<br>$A = P D_A P^{-1} \in P_8$ | $r_1, r_2 \in_R {\mathbb{Z}^*_{251}}^2$<br>$\forall\neq \mu_1 \dots \mu_8 \in_R \mathbb{Z}^*_{251}$<br>$D_B = (\mu_1 \dots \mu_8)$<br>$B = P D_B P^{-1} \in P_8$ |
| | ALICE | BOB |
| Interchange tokens | $A' = A^{k_1} G\, A^{k_2} \Longrightarrow$ | $B' = B^{r_1} G\, B^{r_2} \Longrightarrow$ |
| | ALICE | BOB |
| First common key ($K$) is obtained | $K = A^{k_1} B' A^{k_2}$<br>$m = K^{\nearrow}_{8,1} . K^{\swarrow}_{1,8}$<br>$n = K^{\searrow}_{1,1} . K^{\nwarrow}_{8,8}$ | $K = B^{r_1} A' B^{r_2}$<br>$m = K^{\nearrow}_{8,1} . K^{\swarrow}_{1,8}$<br>$n = K^{\searrow}_{1,1} . K^{\nwarrow}_{8,8}$ |
| | $K = A^{k_1} T_B A^{k_2} = A^{k_1}(B^{r_1} G_0 B^{r_2}) A^{k_2} =$<br>$= B^{r_1}(A^{k_1} G_0 A^{k_2}) B^{r_2} = B^{r_1} T_A\, B^{r_2} = K$ | |

TABLE III
NEW SESSION

| | ALICE | BOB |
|---|---|---|
| *ALICE*<br>start a new cipher session updating recursively parameters | $K = K^{m.n}$<br>$m = K^{\nearrow}_{8,1} . K^{\swarrow}_{1,8}$<br>$n = K^{\searrow}_{1,1} . K^{\nwarrow}_{8,8}$<br>$P = K^m P\, K^n$<br>$G = K^m G\, K^n$<br>$A = P D_A P^{-1}$<br>$A' = A^m G\, A^n \Longrightarrow$ | |

TABLE IV

BOB UPDATES PARAMETERS UPPON ACKNOWLEGMENT.

| | ALICE | BOB |
|---|---|---|
| *BOB*<br>acknowledges and<br>update parameters | | $K = K^{m.n}$<br>$m = K^{\nearrow}_{8,1}.K^{\swarrow}_{1,8}$<br>$n = K^{\searrow}_{1,1}.K^{\nwarrow}_{8,8}$<br>$P = K^m P K^n$<br>$G = K^m G K^n$<br>$B = PD_B P^{-1}$<br>$B' = B^m G B^n \Longrightarrow$ |

TABLE V

ALICE CIPHER AN H MESSAGE TO BOB.

| | ALICE | BOB |
|---|---|---|
| *ALICE*<br>ciphers an *H*<br>message to *BOB* | $H \in M_8$<br>$J \in_R P_8$<br>$C = (y_1, y_2) \Longrightarrow$<br>$y_1 = J^m G J^n$<br>$y_2 = H(J^m B' J^n)$ | |

TABLE VI

BOB DECIPHERS H MESSAGE.

| | ALICE | BOB |
|---|---|---|
| *BOB*<br>deciphers *H* | | $H = y_2(B^m y_1 B^n)^{-1}$ |
| | $H = y_2(B^m y_1 B^n)^{-1}$<br>$= H(J^m B' J^n)(B^m y_1 B^n)^{-1}$<br>$= H(J^m B^m)G_1(B^n J^n)(B^m y_1 B^n)^{-1}$<br>$= H(B^m(J^m G_1 J^n)B^n)(B^m y_1 B^n)^{-1}$<br>$= H(B^m y_1 B^n)(B^m y_1 B^n)^{-1}$<br>$= H$ | |

Suppose that this protocol is intended be used among an *n*-entities community, some caution should be held. The key point would be that each pair of interacting entities should store last interchanged session key until next opened session. That is not a big inconvenience and the protocol remains non-arbitrated.

Another feature could be the incorporation of authentication to block man-in-the-middle attacks. That could be made in a chained mode if each entity begins session exchanging HMAC codes [6] involving the last public key, a timestamp and eventually the last HMAC exchanged.

## 5. STEP-BY-STEP SAMPLE

All symbols used here refers to the previous section. Any interested reader should be able to reconstruct this sequence, as no hidden values are included into this description.

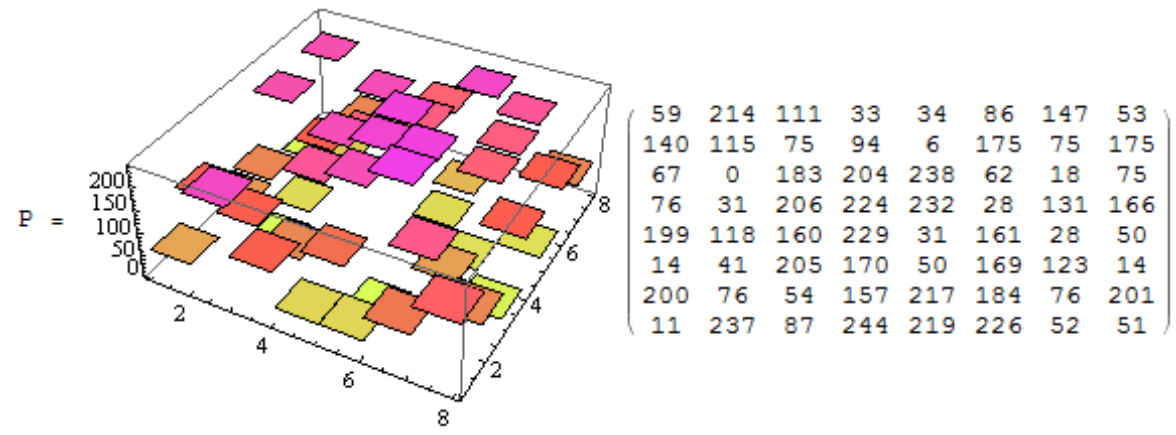

Figure 1. Cipher setup. Any entity defines P and send to the other.

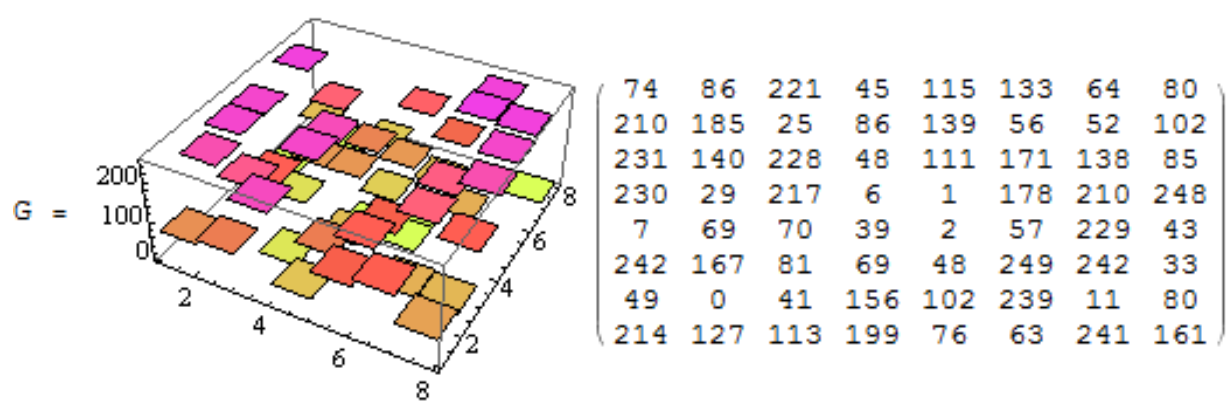

Figure 2. Cipher setup. Any entity defines G and send to the other. It would also be possible that one defines P and the other answers G.

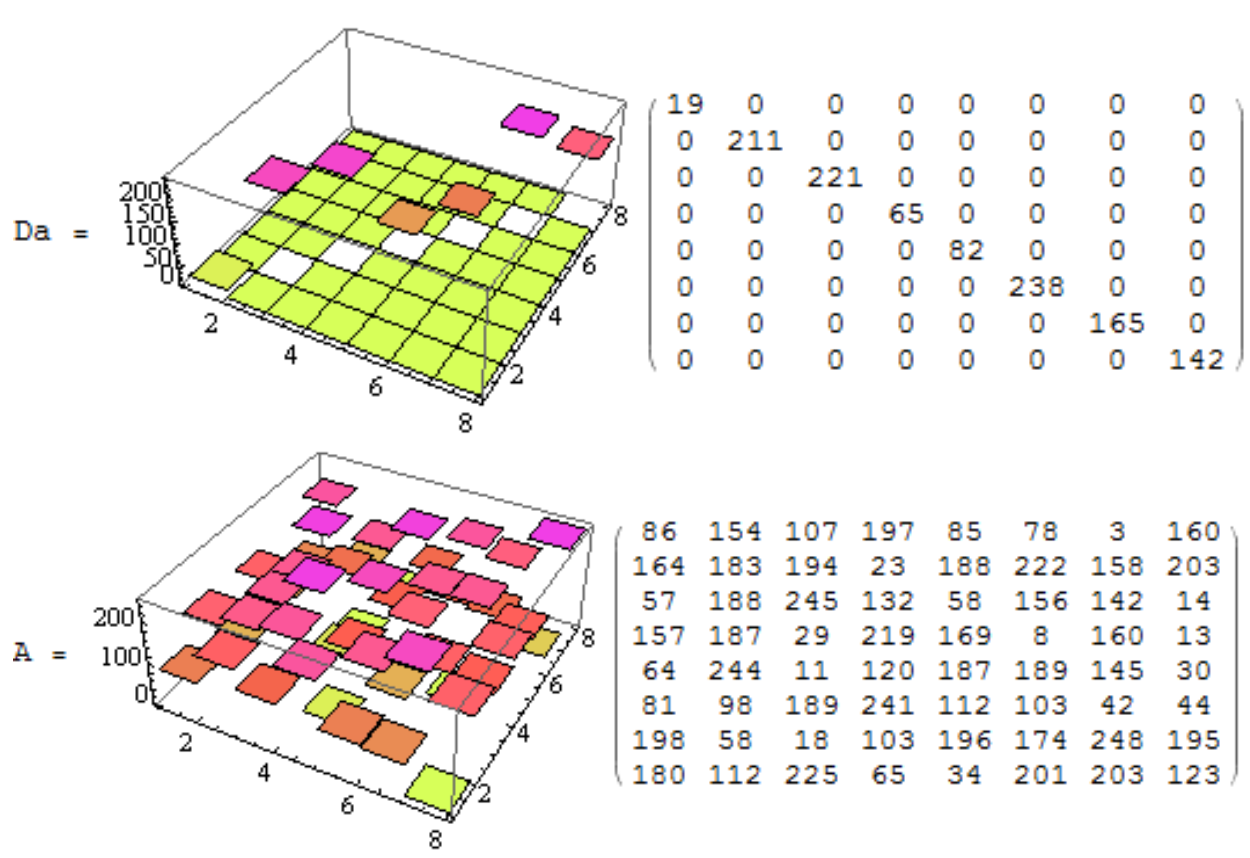

Figure 3. Alice defines her initial private keys. She also has randomly selected $k_1$=77 and $k_2$=184.

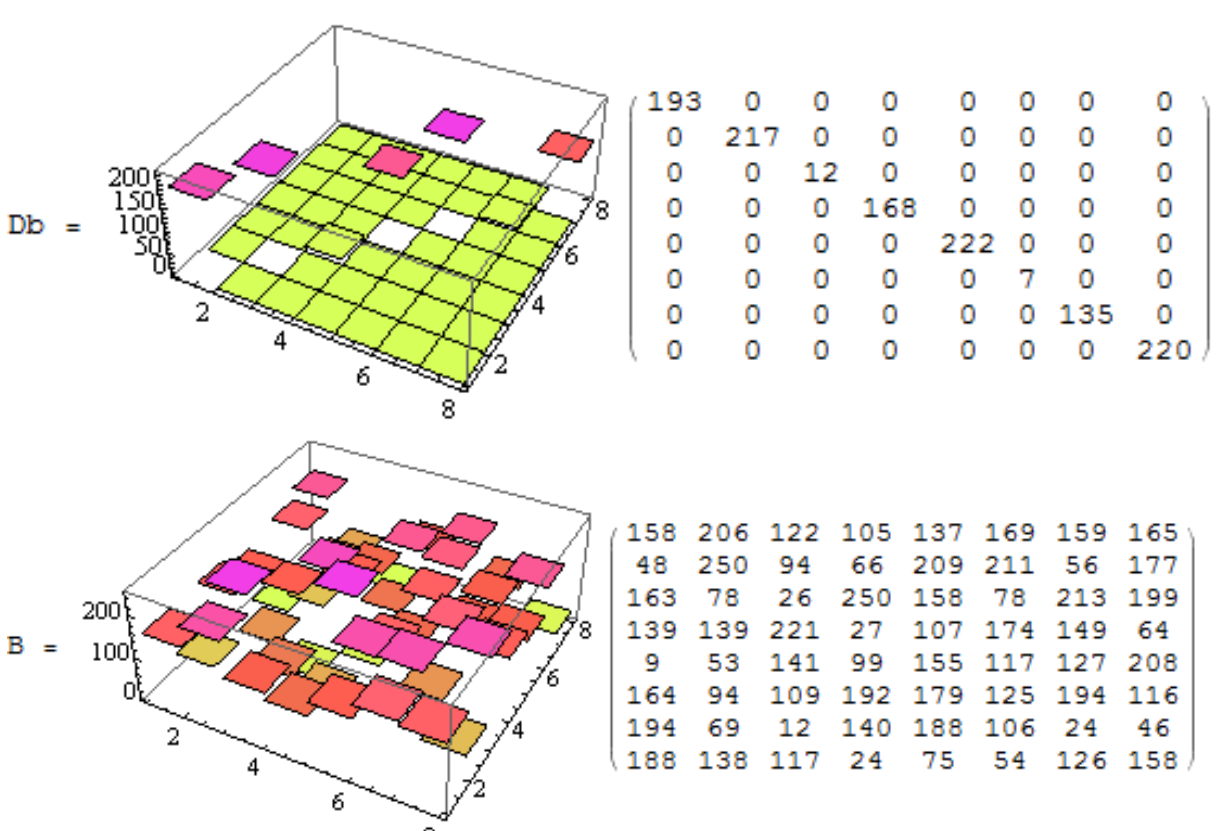

Figure 4. Bob defines his initial private keys. He also has randomly selected $r_1$=42 and $r_2$=229.

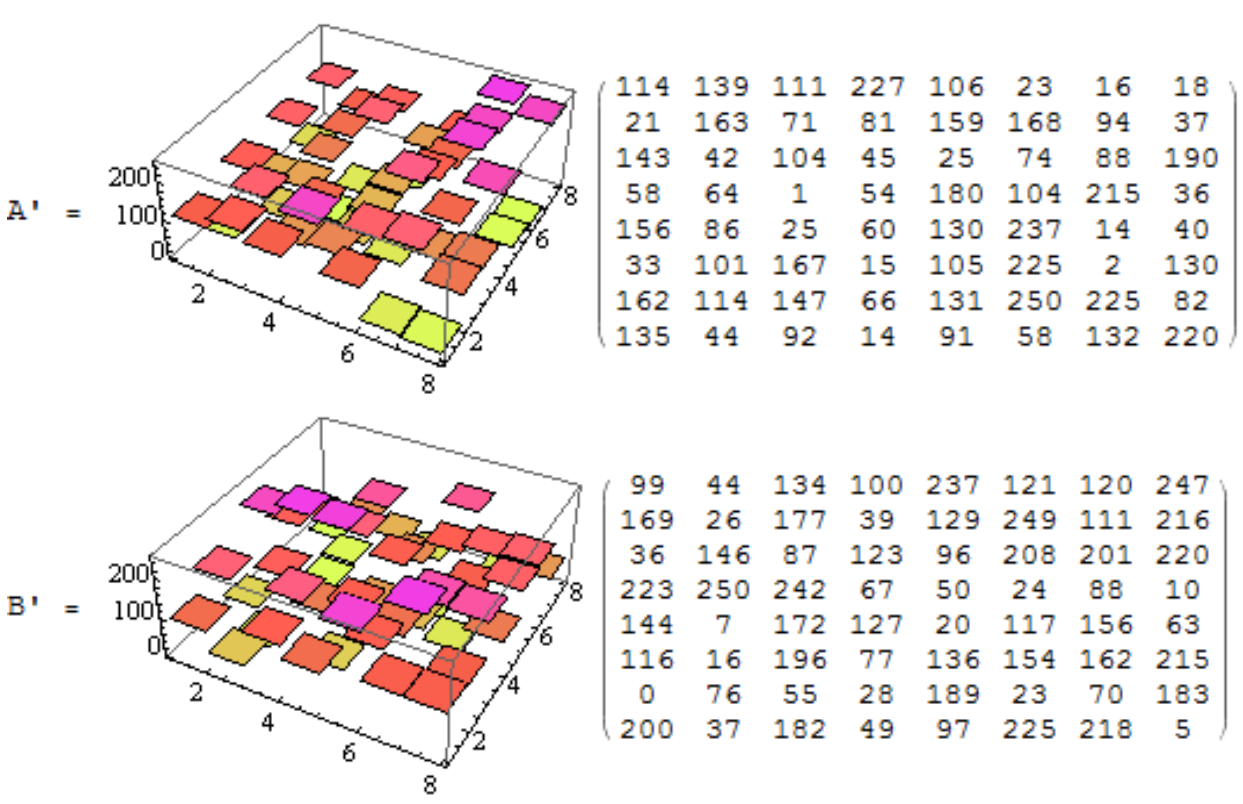

Figure 5. Alice and Bob define tokens and exchange them.

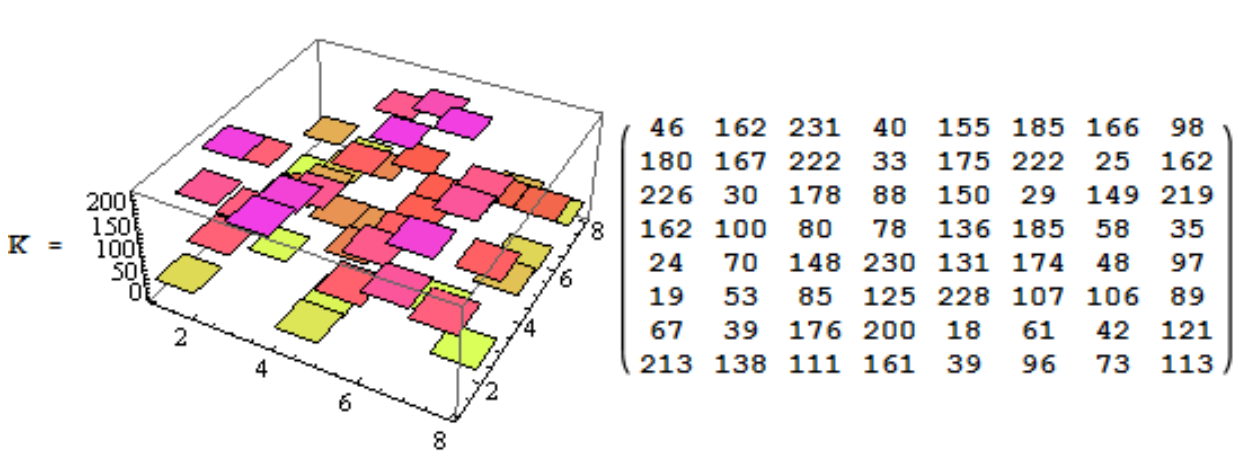

Figure 6. Both obtain the first common session key and the first power parameters using diagonal values ($m$=41, $n$=178, $m.n$= 19).

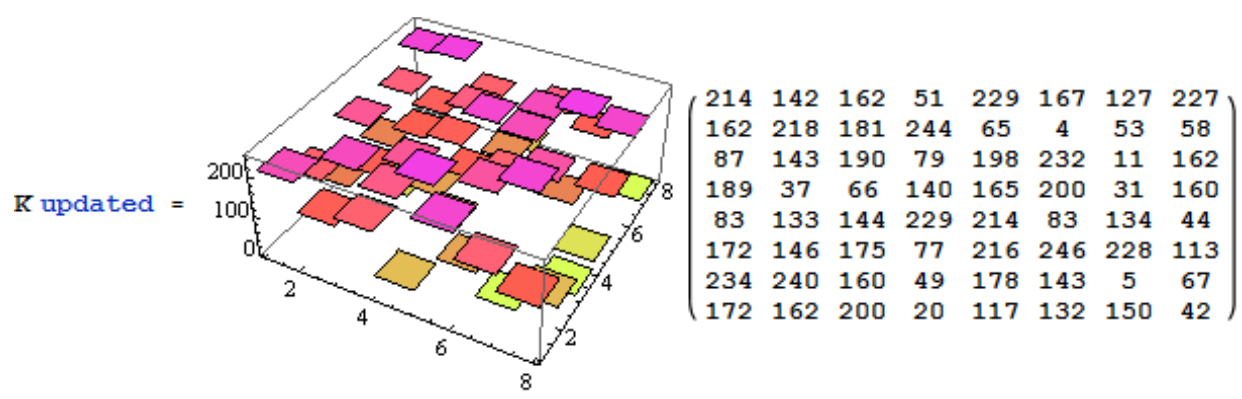

Figure 7. Alice starts a new cipher session. Both update the session key using the current power parameters and calculate new power parameters ($m$=139, $n$=203).

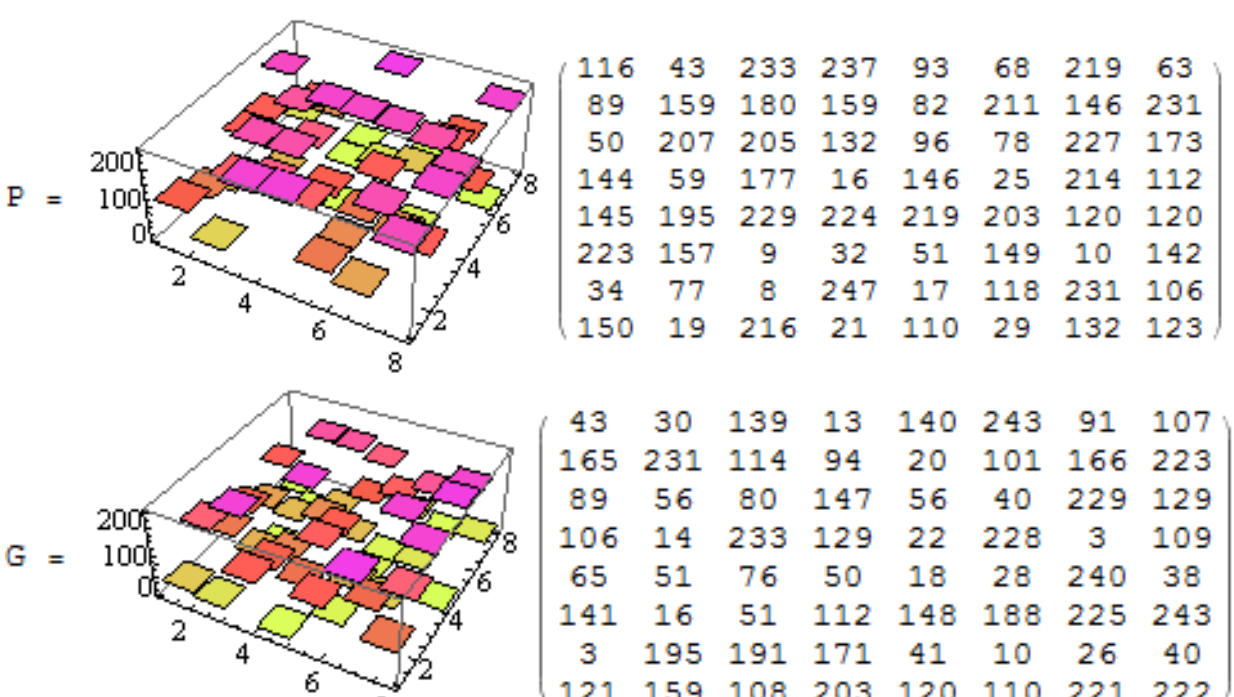

Figure 8. Now both independently update auxiliary matrices.

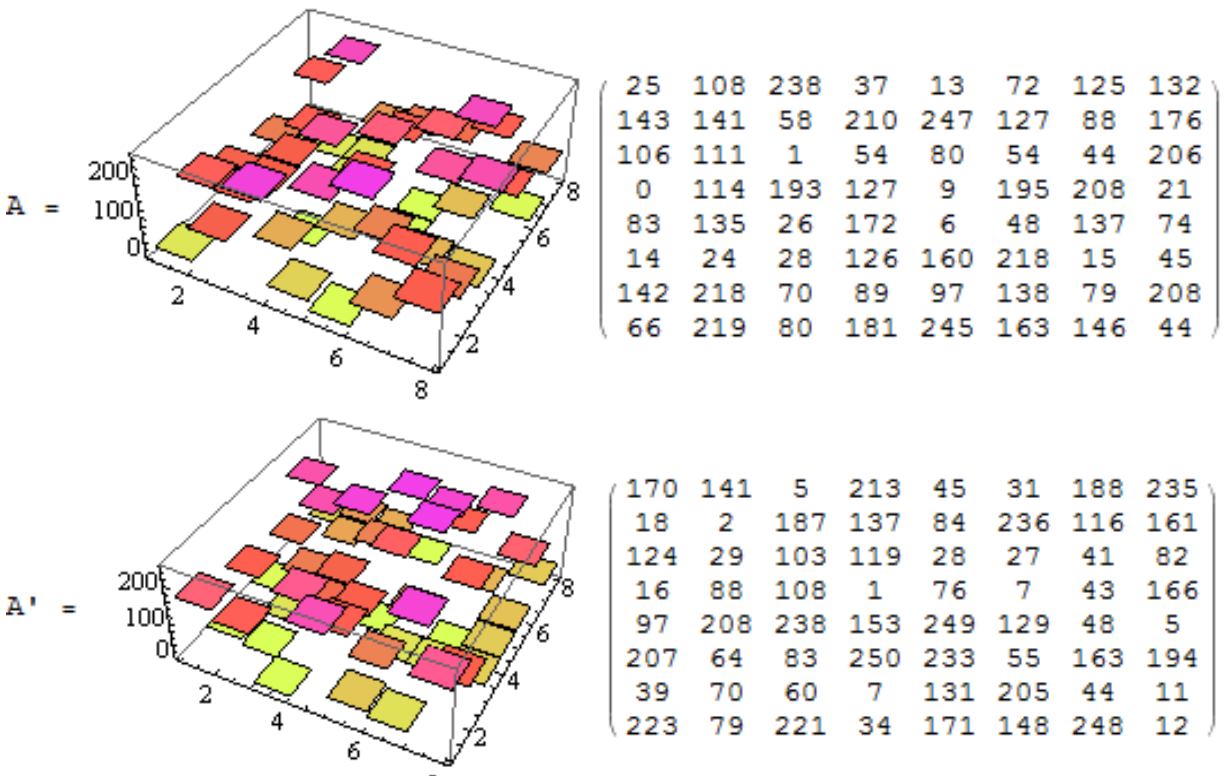

Figure 9. Alice update her private and public session keys. Note that for increased security reason, each new session use recursivelly updated keys.

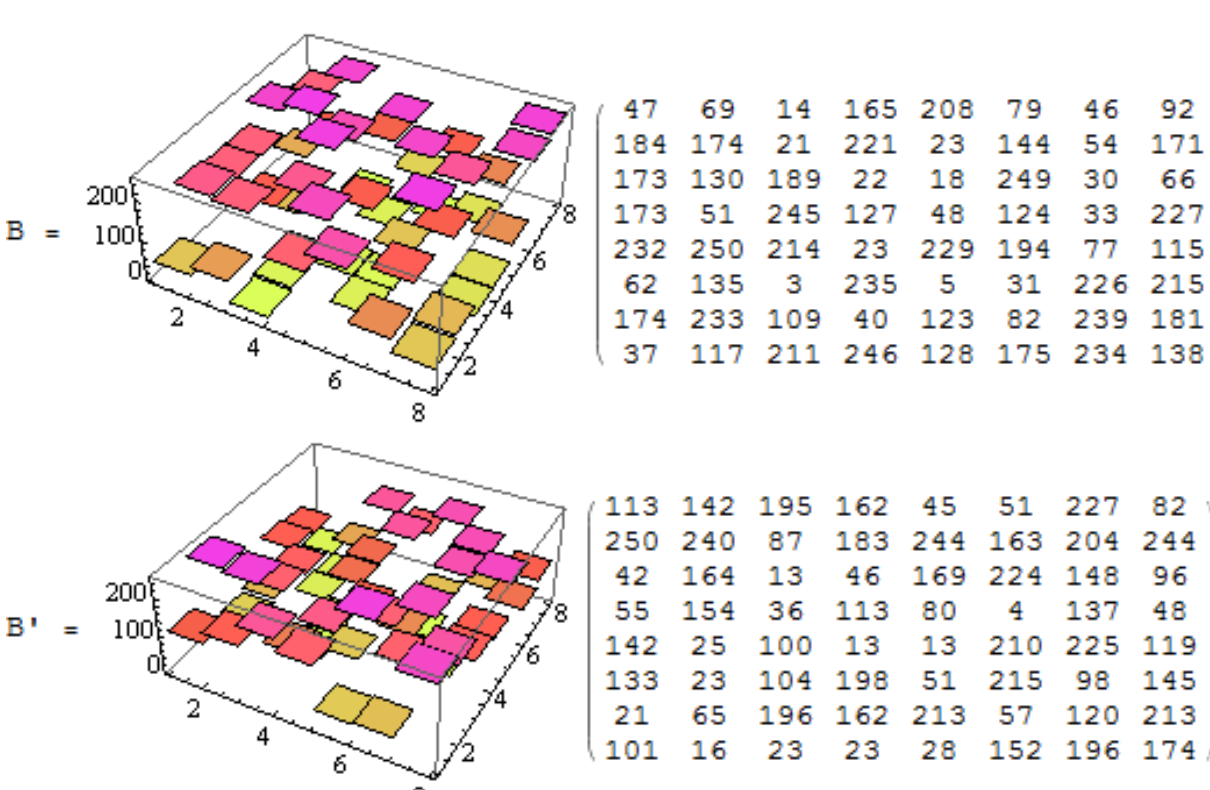

Figure 10. Bob updates his private and public session keys.

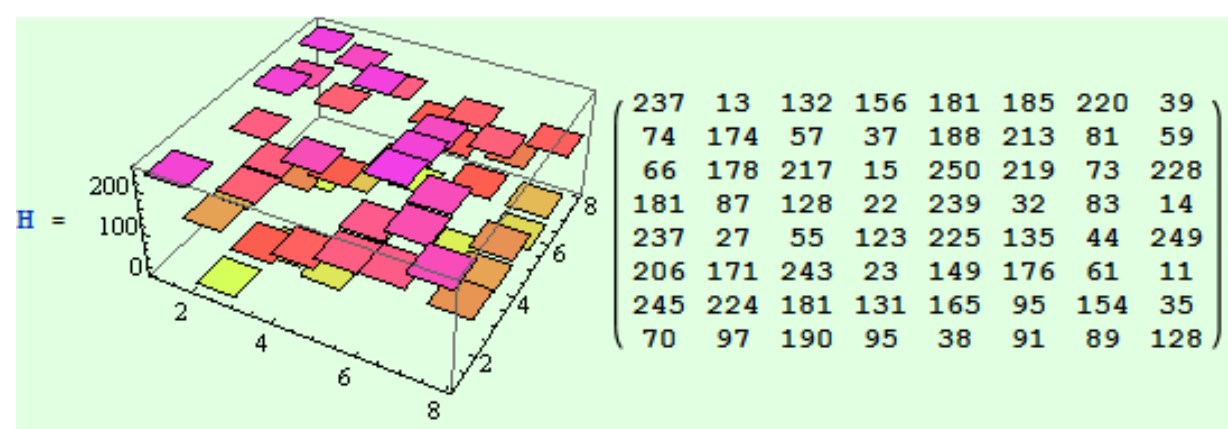

Figure 11. Alice choose H message to cipher. This modular matrix is a general one, the only restriction is to be non-singular.

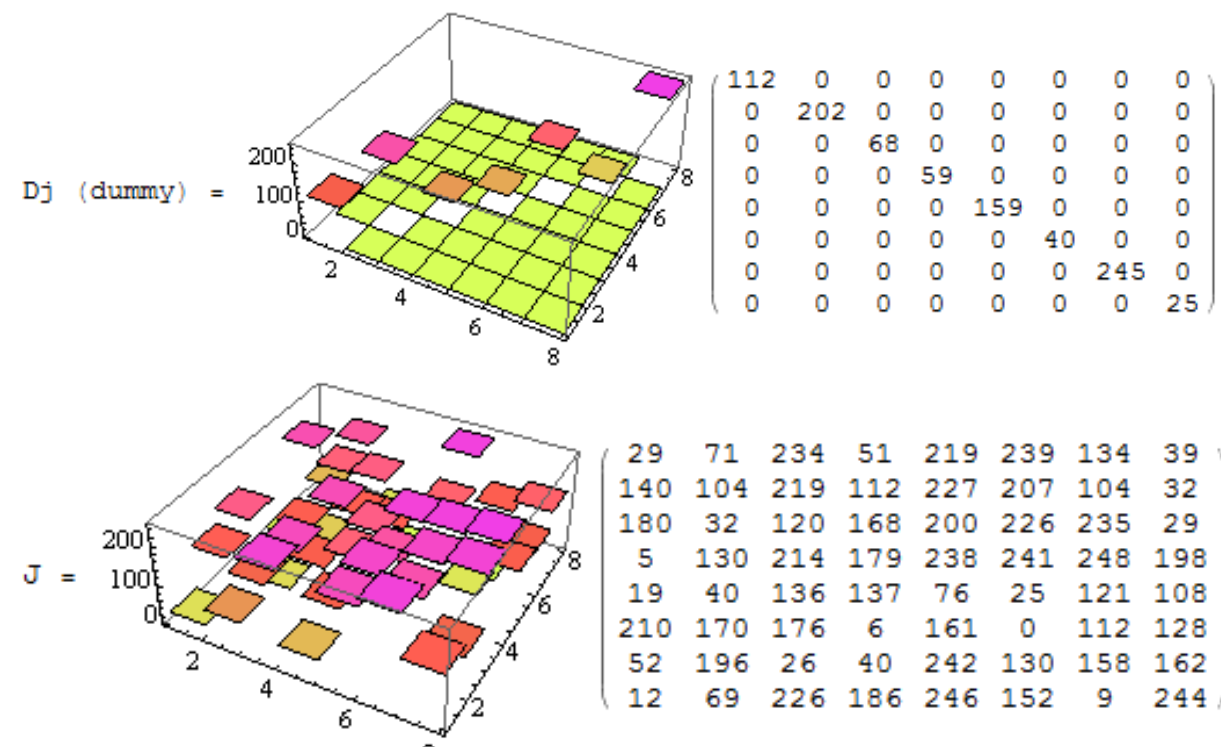

Figure 12. Alice uses a random diagonal matrix to generate a session matrix J. It is mandatory to change J at each cipher session, the same as the k-parameter in a ElGamal numeric field cipher. Please watch out that the updated auxiliary matrix P are used to obtain J.

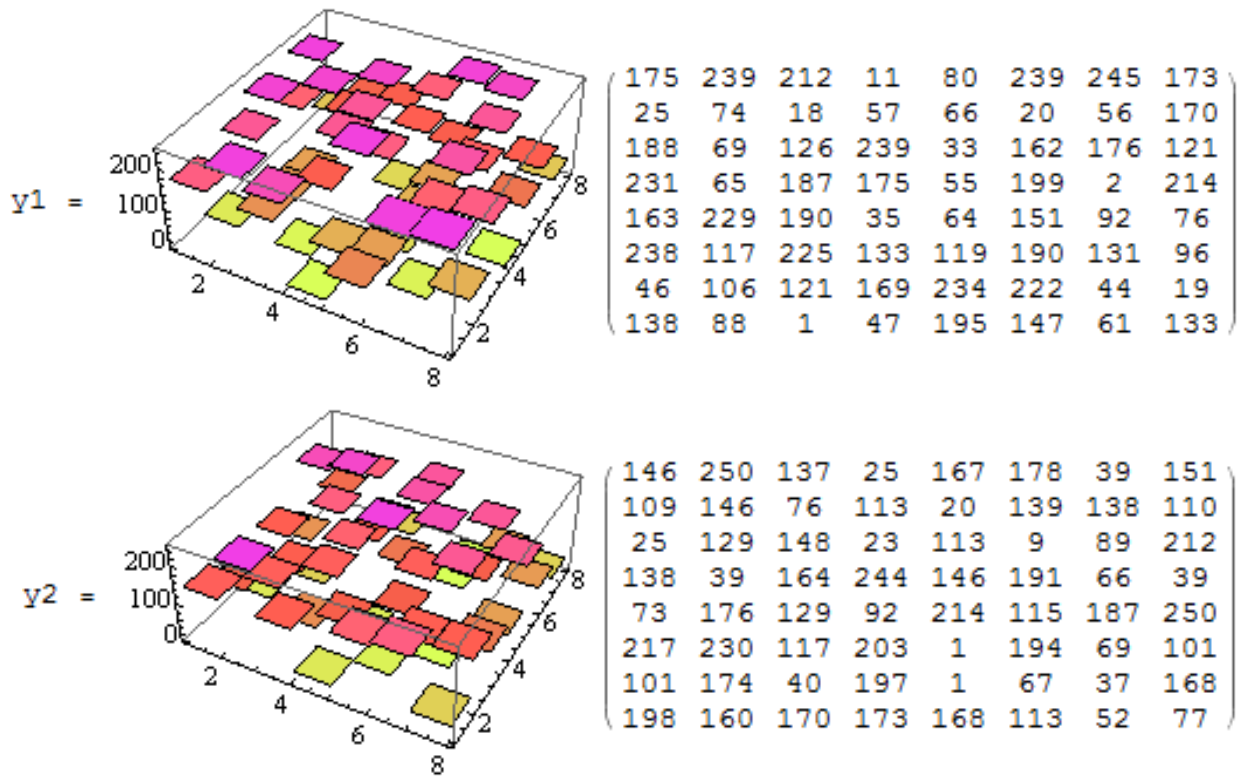

Figure 13. Alice cipher H matrix.

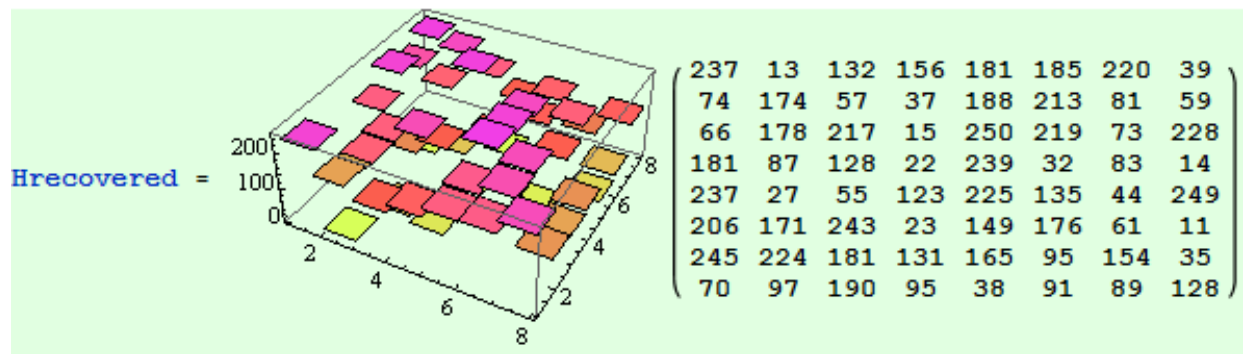

Figure 14. Bob recovers the H message.

## 6. Benchmarking

To estimate the performance of the protocol, we used a simple textbook interpreted program written in *Mathematica 8+* language. This could be one of the worst scenario to test, but it also provides a kind of lower bound for the timing.

The computational platform was an Intel(R) Core(TM) i5-5200U CPU @ 2.20GHz, 2 Core(s), 4 Logical Processor(s) 64-bit Windows 10 Home, version 10.0.14393, 8GB physical RAM in a Dell XPS 13 9343.

The *Mathematica* notebooks here used are freely available upon request. In this simulation sample, instantaneous transfers between entities are assumed, so only computational steps are considered. At same time, no simultaneous or parallel computations are performed, Alice and Bob sum sequentially their timed calculations. All results informed refer to the mean run time of 1000 random iterations.

(a) At setup, definition of $P$ and $G$, took 0.12 ms
(b) From $P$, $G$ already defined until first session key $K$ and new power parameters obtained, took 29.56 ms
(c) New session updating took 52.94 ms
(d) Enciphering–deciphering cycle took 32.36 ms

As observed, a full session of an approximate 64-bytes message (an $H$ matrix) secured transmission took 85 ms in our environment. Of course, a lot of optimization should be accomplished before a real-life application is planned.

## 7. Protocol Security

The group of order 8 modular integer matrices $GL(8, F_{251})$ has a cardinal $251^{64} \approx 10^{153.579}$. The invertible Hill matrices subgroup $M_8$=$GL(8, F_{251})$ has a slighty lower order [36]

$$251^{64}(1\text{-}1/251)(1-1/251^{2})(1-1/251^{3})(1-1/251^{4})(1\text{-}1/251^{5})(1-1/251^{6})(1-1/251^{7})(1-1/251^{8}) \approx 10^{153.177} \quad (6)$$

Comparing both numbers, the probability of selecting a singular matrix in $M(8, Z_{251})$ is $p \approx 0.004$, a low but not negligible value. Each time a new random modular matrix is obtained, it must be controlled that his determinant is not null.

Supposing no other weakness are available, cracking a private key depends on an order eight diagonal matrix, so a brute force search of the commutative $P_8$ subgroup of $M_8$ involves the cardinal

$$|P_8| = 249.248.247.246.245.244.243.242 = \\ = 13190481178699144320 \approx 10^{19} \approx 2^{64} \quad (7)$$

Currently it is impossible to make a systematic search of that space, and if a greater security is pursued, it would suffice to expand the commutative subgroup to $P_{16}$, who implies a 127-bit level. It is recommended to adopt a compromise solution between the desire to obtain greater security and the concomitant use of more resources, which are always costly and limited.

A second way to attack the present protocol would be to find a polynomial time algorithm to solve the algebraic generalized symmetric decomposition. As some simpler OWTF based on algebraic conjugation were successfully cryptanalyzed [38, 39], it was mandatory to find very hard functions. We presented earlier (see definition 5.) a stronger version, the blind general symmetric decomposition problem (BGSDP). As posted, it could be conjectured that this kind of algebraic challenge belongs to a *NP* time-complexity class and at same time resilient to quantum computers attacks. As said, this statement is currently unproved and it seems not easy to be solved.

Perhaps there exists a completely different way to attack the present protocol; but at current time the author is unaware of it. As consequence, we assume a 64-bit security for the protocol as it is stated.

## 8. Conclusions

We developed a non-arbitrated and compact algebraic post-quantum cipher protocol, which could easily be adapted to other purposes as key exchange, key transport and ZKP authentication [9, 30]. By compact, we mean that no big number library is required as only $Z_{251}$ field operations are involved. This feature would enable the use or it in low computational resources environments like smartphones, smartcards, etc.

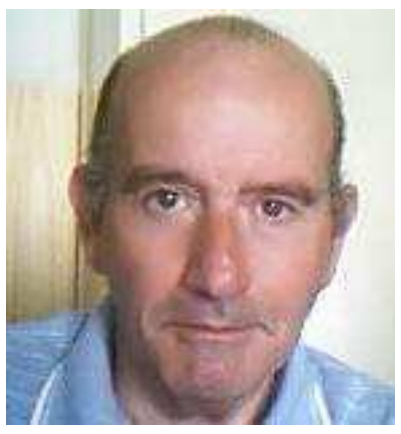

**Pedro Hecht** received an MSci in Information Technology at Escuela Superior de Investigación Operativa (ESIO-DIGID) and an PhD degree from Universidad Nacional de Buenos Aires (UBA). Currently, he is full professor of cryptography at Information Security Graduate School at UBA, EST (Army Engineering School) and IUPFA (Federal Police University), he is also a research fellow UBACyT and Director of EUDEBA editorial board of UBA. Current fields of interest are algebraic PQC solutions like non-commutative, non-associative structures.